\documentclass[%
 aip,
rsi,%
 amsmath,amssymb,
 reprint,%
]{revtex4-1}
\usepackage{subfigure}
\usepackage{graphicx}% Include figure files
\usepackage{dcolumn}% Align table columns on decimal point
\usepackage{bm}% bold math
\usepackage{todonotes}
\begin{document}

%\preprint{AIP/123-QED}

\title[Hong--Ou--Mandel-like two-droplet correlations]{Hong--Ou--Mandel-like two-droplet correlations}% Force line breaks with \\

\author{Rahil N. Valani}
\email{Rahil.Valani@monash.edu}
\affiliation{School of Physics and Astronomy, Monash University, VIC 3800, Australia}
\author{Anja C. Slim}%
\affiliation{School of Mathematical Sciences, Monash University, VIC 3800, Australia}
\affiliation{School of Earth, Atmosphere and Environment, Monash University, VIC 3800, Australia}
\author{Tapio Simula}
\affiliation{School of Physics and Astronomy, Monash University, VIC 3800, Australia}

\newcommand{\RV}[1]{{\protect \todo[inline, color =orange]{RV---#1}}}
\newcommand{\AS}[1]{{\protect \todo[inline, color =yellow]{AS---#1}}}
\newcommand{\TS}[1]{{\protect \todo[inline, color =gray]{TS---#1}}}
\newcommand{\JCM}[1]{{\protect \todo[inline, color=green]{JCM---#1}}}

\begin{abstract}
We present a numerical study of two-droplet pair correlations for in-phase droplets walking on a vibrating bath. Two such walkers are launched towards a common origin. As they approach, their carrier waves may overlap and the droplets have a non-zero probability of forming a two-droplet bound state. The likelihood of such pairing is quantified by measuring the probability of finding the droplets in a bound state at late times. Three generic types of two-droplet correlations are observed: promenading, orbiting and chasing pair of walkers. For certain parameters, the droplets may become correlated for certain initial path differences and remain uncorrelated for others, while in
other cases the droplets may never produce droplet pairs. These observations pave the way for further studies of strongly correlated many-droplet behaviors in the hydrodynamical quantum analogs of bouncing and walking droplets.
\end{abstract}

\maketitle
 
\begin{quotation}
Hong-Ou-Mandel (HOM) interference\citep{PhysRevLett.59.2044} is a well-known phenomenon in quantum optics, in which two photons pass through a beam splitter before being observed at two photodetectors. By counting coincident photon arrivals in the two detectors, it is possible to reveal two-photon correlations and quantum entanglement that are only possible when the photons are indistinguishable. Inspired by this phenomenon, we numerically investigate emergent two-droplet correlations of walking droplets on a vibrating fluid surface. We launch pairs of droplets, initially separated by different distances away from the intersection of the lines defined by their initial velocity vectors, towards a common region of space. For certain initial path differences the droplets are observed to develop non-trivial correlations, binding into two-droplet dimers mediated by the interference of their carrier waves. These results open the pathway for studies of hydrodynamical quantum analogs of the HOM phenomenon.%\AS{I think there should be a reference in here for HOM}\TS{Quick look at other Chaos papers seems to indicate that the bold paragraph should be unreferenced?}\AS{The instructions say it can be referenced.}
\end{quotation}

\section{Introduction}

On vibrating a bath of silicone oil vertically with frequency $f$ and time-dependent acceleration $a(t)=\gamma \cos(2\pi f t)$, a droplet of the same liquid can be made to bounce indefinitely provided that $\gamma>\gamma_{\rm B}$, where $\gamma_{\rm B}$ is the bouncing threshold.\citep{PhysRevLett.94.177801} If the peak forcing acceleration $\gamma$ exceeds the Faraday threshold value $\gamma_{\rm F}$, then the whole liquid-air interface becomes unstable and standing surface waves emerge that oscillate at a frequency $f/2$. In between the bouncing threshold and the Faraday threshold, for certain size droplets, a period doubling bifurcation takes place and the bouncing state destabilizes into a robust walking state for $\gamma_{\rm W}<\gamma<\gamma_{\rm F}$, where $\gamma_{\rm W}>\gamma_{\rm B}$ is the walking threshold.\citep{moláček_bush_2013,Molacek2013DropsTheory} In this walking state, the droplet bounces at frequency $f/2$ and is a local exciter of Faraday waves that in turn propel the droplet through resonant interaction with the wavefield. Consequently the bouncing droplet is a realization of a discrete time crystal. \citep{Zhang2017}%\AS{I don't understand this term and so the relevance of this sentence}
%\TS{Check out the Wikipedia page of ``Time Crystal"...}\AS{I'm still not clear on what the relevance is.}\TS{The relevance is the fact that it is a fact :)
%
%1. Two papers were published in Nature only last? year reporting the first experimental realizations of discrete time crystals which are not much different from ours at all!
%2. This is an emerging hot topic and is drawing a lot of attention at the moment.
%3. This creates a hook/link to this fact that we and others can cite from later publications.
%4. One of the goals of this manuscript is to draw interdisciplinary attention bridging ``droplets and quantum" communities but hopefully even more broadly.
%5. It is one more special property pertinent to the droplet system.
%6. list goes on...}
%\AS{I guess I feel it would be useful to give more context to an audience who are not familiar with the term.  Are there insights from time crystals that would be useful to droplets or vice versa?  Do droplets provide a particularly realisable time crystal or one with particular properties that would allow easy insight into some of the questions?  Basically is there something short and powerful that could be said here eg. "these have many of the features of spatial crystals, including ... but don't do ... and suggest that the system will exhibit ... or will provide insight into ..."?  Rahil - if this is easy and you feel comfortable with it, maybe do it, if not just leave as is.}

\begin{figure*}
\centering
\includegraphics[width=1.8\columnwidth]{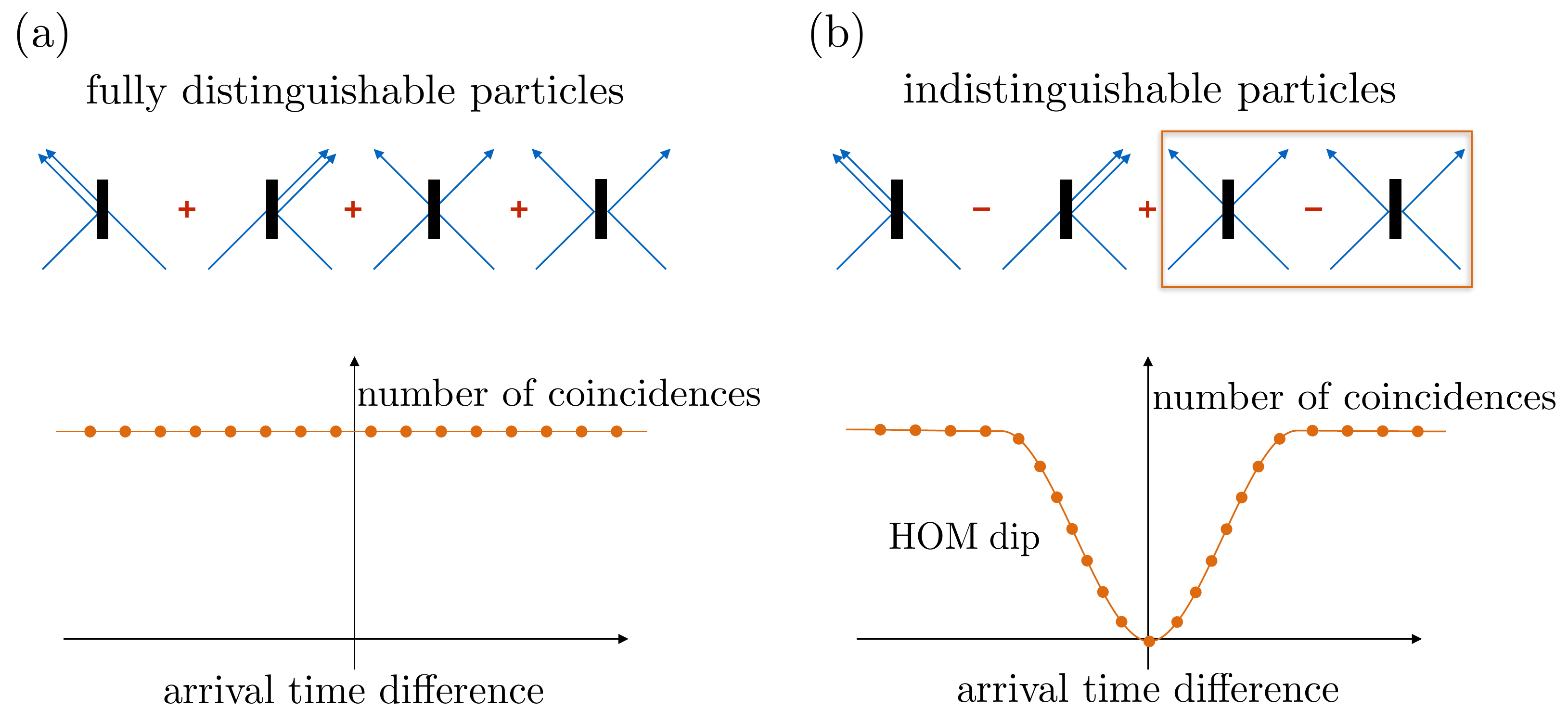}
\caption{Hong--Ou--Mandel interference of photons. (a) Two distinquishable photons (blue arrows) may be combined in a beam splitter (black bar) in four ways with equal probabilities and the probability of coincident detection (one photon leaving on each side of the beam splitter) is independent of the delay in the time of arrival of the photons at the beam splitter. (b) Two indistinquishable photons may also be combined in a beam splitter in four ways but the quantum amplitudes of `neither is reflected' and `both are reflected' interfere destructively such that no coincident detection is possible when the photon arrival time difference vanishes. This results in the Hong--Ou--Mandel dip in the coincidence detection probability as a function of the time of arrival of the photons at the beam splitter. }
\label{schematicHOM}
\end{figure*}

Just below the Faraday threshold, the waves generated by the walking droplet during each bounce are long lived and thus the droplet is not only affected by the wave from its previous impact, but also from the waves it generated in the distant past---the droplet has a memory and its behaviour is characterised by non-Markovian dynamics. In this high memory regime where $\gamma \to \gamma_{\rm F}^-$, walking droplets have been shown to mimic several quantum phenomena. Although claims of single particle diffraction through single and double slit arrangements \citep{Couder2006} have been contested, \citep{PhysRevE.92.013006,pucci_harris_faria_bush_2018} hydrodynamical quantum analogs such as orbital quantization in a rotating frame\citep{Oza2014} and in a harmonic potential,\citep{Perrard2014a,Perrard2014b} tunneling across submerged barriers\citep{Eddi2009} and wavelike statistics in confined geometries \citep{PhysRevE.88.011001,Sáenz2017} are robust phenomena. 

Two walkers interact with each other through their wavefields and exhibit intricate dynamics. Two identical walkers can either scatter or bind into states such as parallel walkers, orbiting pairs and promenading pairs (parallel walkers that oscillate towards and away from one another while walking)\cite{protière_boudaoud_couder_2006} while mis-matched walkers can also form ratcheting pairs.\citep{Couder2008} Multiple bouncing droplets have been shown to self-organize into lattice structures\cite{protière_boudaoud_couder_2006} and form drifting rafts.\citep{Couder2008} \citet{Borghesi2014} studied promenading pairs both theoretically and experimentally and related the energy stored in the wavefield to the interaction between the walkers. \citet{PhysRevE.78.036204} experimentally investigated the orbiting dynamics of two droplets with different sizes and found circular, oscillating and epicycloidal orbits. Recently, both orbiting\citep{PhysRevFluids.2.053601} and promenading pairs\citep{PhysRevFluids.3.013604} have been revisited theoretically and experimentally and it was shown that the impact phase of walkers vary in these bound states and that that variation plays an important role in stabilizing these states. So far the impact phase has only been accounted for empirically and a complete theoretical description characterising the time-dependent behaviour of the impact phase is lacking. Generically, the non-Markovian dynamics of the droplets, coupled with many-droplet interactions, results in rich correlated behaviors in these system. 

In quantum mechanical systems, particle correlations are of fundamental importance. The Einstein--Podolsky--Rosen paradox\citep{PhysRev.47.777} and the Hanbury Brown and Twiss effect\citep{HANBURYBROWN1956} are vivid demonstrations of non-classical correlations. Quantum correlations that have no classical counterpart can also be revealed using the Hong--Ou--Mandel (HOM) two-photon interference experiment.\citep{PhysRevLett.59.2044} In the classic optical HOM effect, illustrated in Fig.~\ref{schematicHOM}, two photons (blue arrows) arrive at a `50/50' beam splitter. A single photon, when incident on such a beam splitter, has a 50\% probability of being reflected and a 50\% probability of being transmitted. When two photons are incident, four possibilities arise: (1) the photon coming from the left is reflected and the photon arriving from the right is transmitted, (2) the photon coming from the right is reflected and the photon arriving from the left is transmitted, (3) both photons are transmitted, and (4) both photons are reflected. Two detectors (not shown) are placed far behind the beam splitter, one on the left and the other on the right side, and record coincident photon pairs (one photon detected by each detector). The normalized number of coincidence detections are recorded as a function of the difference in time of arrival of the photons. 

If the photons are fully distinguishable, Fig.~\ref{schematicHOM}(a), all four possibilities occur with equal probabilities and whether the photons pass through the beam splitter simultaneously or one after another is irrelevant. However, if the photons are indistinguishable, Fig.~\ref{schematicHOM}(b), their quantum mechanical description shows that the last two of the four possible outcomes cancel out---the photons interfere destructively---and the probability of coincident detection vanishes. One way to continuously `tune the level of indistinguishably' is to vary the distance of the photon sources from the detector and thereby the difference in the photon arrival times. This results in the HOM \emph{dip} shown in Fig.~\ref{schematicHOM}(b) whose characteristic width is determined by the size of the wavepacket of the photons. The HOM effect has a classical analog and a HOM dip is also observable using classical light sources. However, the maximum visibility of the HOM dip cannot exceed $0.5$ (the coincidence detection rate at the deepest point of the dip cannot be less than $0.5$ of the maximum coincidence detection rate) for classical waves or particles.\cite{PhysRevLett.96.240502} As such, the HOM effect can be used for drawing a distinction between classical and quantum correlations. More recently, the HOM interference has been observed for photons that always pass through the beam splitter at different times\cite{Kim2016} and by using atoms instead of photons.\citep{Lopes2015} Also, a variant of the HOM experiment in the absence of beam splitters has been proposed.\cite{Mährlein2017}

Inspired by the atomic and optical HOM phenomena, we have performed numerical experiments with two walkers to study their spatio-temporal correlations. Nevertheless, we emphasize that, unlike photons, our droplets are interacting particles and cannot be indistinguishable in the quantum mechanical sense. A simple thought experiment is sufficient to demonstrate this: filling one of the droplets with dye will facilitate tracking the exact paths of the two droplets yet the results presented here would be unaffected by such particle tagging.

The article is organised as follows. In Section II we present the numerical model and methods used to simulate the non-Markovian dynamics of the droplets. Two-droplet correlations are discussed in Section III. The conclusions are provided in Section IV.

\section{Model}
\subsection{Equations of motion}
Consider two identical droplets of mass $m$ and diameter $d$ bouncing at frequency $f/2$ on the surface of a bath that is oscillating vertically at frequency $f$. {\color{black}In this work we only consider in-phase interactions of the two droplets}. The positions (in units of $k_F^{-1}=\lambda_{\rm F}/2\pi$, where $\lambda_{\rm F}$ is the Faraday wavelength) of the two droplets in the horizontal plane are given by $\mathbf{r}_1=(x_1,y_1)$ and $\mathbf{r}_2=(x_2,y_2)$. We describe the horizontal motion of the droplets using the model of Oza \textit{et al.},\citep{PhysRevFluids.2.053601} a stroboscopic approximation for two-droplet dynamics. In this model the impact phase is adjustable based on empirical observations;{\color{black}\citep{PhysRevFluids.2.053601,PhysRevFluids.3.013604}} in our theoretical exploration we set it to be a representative constant. The dimensionless equations of motion for two walkers are thus
\begin{align}\label{eq_1}
\kappa\frac{d^2{\bf r}_i}{dt^2}+\frac{d{\bf r}_i}{dt}
={\color{black}-}\beta \nabla h(\mathbf{r},t)\big|_{\mathbf{r}_i} \text{ for } i=1, 2,
\end{align}
with wavefield
 \begin{align*}\label{h_eq}
h({\bf r},t) &= \int_{-\infty}^{t}\text{J}_0(|\mathbf{r} - \mathbf{r}_{1}(s)|)\,\text{e}^{-G_1({\bf r},t,s)}\text{d}s \notag\\
& + \int_{-\infty}^{t}{\text{J}_0(|\mathbf{r} - \mathbf{r}_{2}(s)|)}\,\text{e}^{-G_2({\bf r},t,s)}\text{d}s
\end{align*}
and spatial and temporal decay envelope
\begin{equation*}
G_j({\bf r},t,s)=\hat{\alpha}{\frac{ [\mathbf{r} - \mathbf{r}_{{\color{black}j}}(s)]^2}{t-s+{M_e}^{-1}}+(t-s)} \text{ for } j=1, 2.
\end{equation*}
The dimensionless parameters $\kappa$, $M_e$ and $\beta$ follow directly from the stroboscopic model for a single walker developed by \citet{Oza2013} and are given by $\kappa=m/D T_F M_e$, $\beta=mgAk_F^2 T_F M_e^2/D$ and $M_e=T_d/T_F(1-\gamma/\gamma_F)$, while the parameter $\hat{\alpha}=\alpha/k_F^2 T_F M_e$ follows from \citet{PhysRevFluids.2.053601} Here $T_F=2/f$ is the Faraday period and $T_d$ is the decay time of waves in the absence of forcing. Here time has been non-dimensionalized using the scale $T_F M_e$. The parameters $A$ and $D$ are the wave amplitude and the drag-force coefficient respectively; we refer the reader to \citet{Oza2013} for their complete dependence on system parameters. The first term in Eq.~(\ref{eq_1}) is an inertial term with an effective mass $\kappa$ and the second term models the average drag over a bouncing period consisting of aerodynamic drag during flight and {\color{black}drag during} momentum transfer between the droplet and the bath during impact.\citep{Oza2013} 
 
The wave generated by each impact of a droplet is modelled as an axisymmetric Bessel function $\text{J}_0(\mathbf{r})$ with a Gaussian envelope centered at the point of impact. This standing wave decays exponentially in time. Since this model accounts for the waves generated from all the previous impacts, the shape of the interface is calculated through integration of waves generated from all the previous bounces of both droplets. At each impact, the droplet receives a horizontal kick proportional to the gradient of the liquid-air interface $h({\bf r},t)$ evaluated at that point. The strength of this non-Markovian force is modulated by the $\beta$ parameter.

\subsection{System parameters}
We restrict our exploration of the parameter space by fixing the parameters to the typical values for experiments\citep{Oza2013,PhysRevFluids.2.053601,PhysRevFluids.3.013604} {\color{black} in addition to limiting to in-phase droplets}. We consider a fixed forcing frequency of $f=80\;$Hz and consider droplets of diameter $d$ in the range $0.6\;\text{mm}\leq d \leq 1\;\text{mm}$ that are typically found in experiments.\citep{Bush2015} In accordance with the recent experiments on orbiting\citep{PhysRevFluids.2.053601} and promenading\citep{PhysRevFluids.3.013604} pairs, the fluid density is chosen to be $\rho=949\;$kg/m$^3$, fluid viscosity $\nu=20$\;cS, surface tension $\sigma=20.6\times 10^{-3}$\;N/m, $\lambda_F=4.75$\;mm, $\gamma_F=4.2$g with g being the gravitational acceleration, $T_d=1/54.9$\;s, viscosity of air $\mu_a=1.84 \times 10^{-5}$\;kg/ms and density of air $\rho_a=1.2$\;kg/m$^3$. We choose a constant impact phase of $\sin\Phi=0.2$. The dimensionless drag coefficient $C$ depends weakly on system parameters\citep{Molacek2013DropsTheory} and is shown to vary over the range $0.17 \leq C \leq 0.33$. In our numerical experiments we consider the two extreme values $C=0.17$ and $0.33$. 
\newline
\newline
{\color{black} From here on, we will use dimensionless quantities in the results with the length scale chosen as the Faraday wavelength $\lambda_F$.}  
\begin{figure}[t]
\centering
\includegraphics[width=\columnwidth]{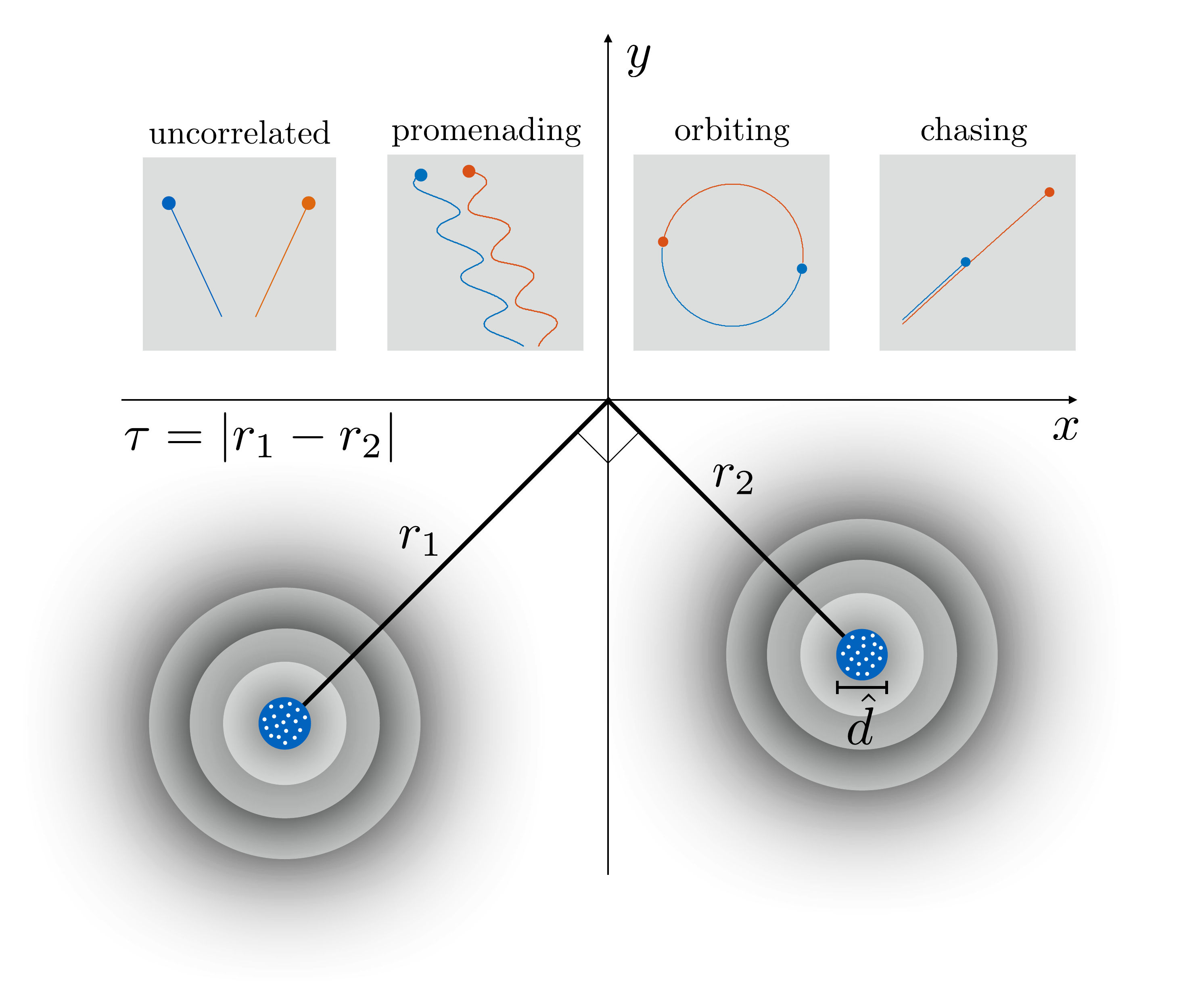}
\caption{Schematic diagram of the numerical experiments. Each droplet has a {\color{black}dimensionless} diameter $\hat{d}$. The initial path difference is denoted by $\tau$. The initial center-of-mass positions (white dots) of the droplets are randomly chosen withing a disk of diameter $\hat{d}$. The initial velocities have constant magnitude $V$ and are always oriented toward the origin. Four generic types of behavior are observed at late times (top sketches): (i) uncorrelated, (ii) {\color{black}promenading} correlations, (iii) orbiting correlations, and (iv) chasing correlations.}
\label{schematic}
\end{figure}
\begin{figure*}[t]
\centering
\includegraphics[width=1.5\columnwidth]{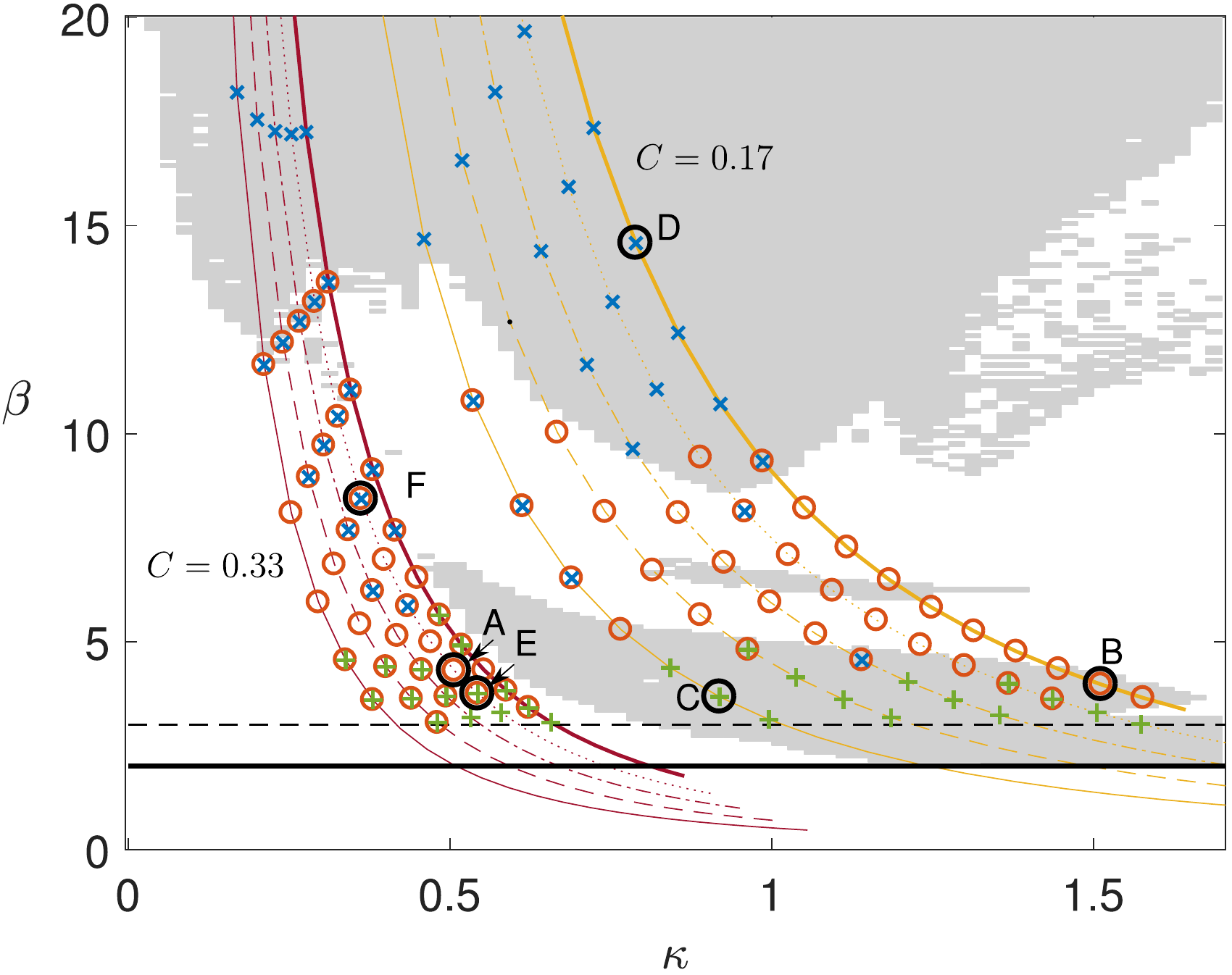}
\caption{The $\beta$-$\kappa$ parameter space considered in the numerical experiments. The two sets of curves correspond to dimensionless drag coefficients $C=0.17$ (yellow) and $C=0.33$ (dark red). Each of the five curves of the same color corresponds to droplets of diameters $d=0.6$ (solid), $0.7$ (dashed), $0.8$ (dashed-dotted), $0.9$ (dotted) and $1\;$mm (thick solid). Light gray shaded regions correspond to parallel walkers that unbind at long times in simulations with $\hat{\alpha}=0$, while white regions correspond to parallel walkers that remain as stable droplet pairs at long times, see Valani and Slim.\citep{twodroplets} The black horizontal line indicates the walking threshold $\beta=2$. All simulations were performed for $\beta\geq 3$ (black dashed line) to ensure that the walking speed of the droplet is sufficiently high so that a single walker would travel at least $2r_1$ by the end of the simulation, $t=250$. The markers indicate the three different types of correlations observed: {\color{black}promenading} (red $\circ$), orbiting (green $+$) and chasing (blue $\times$) {\color{black}pairs of walkers}.  {\color{black}Simulations with only uncorrelated walkers are shown as a black dot.} {\color{black} Markers indicate the correlations observed with probability greater than 20\% for simulations with path differences $0\leq \tau \leq 18/2\pi$.} 
}
\label{ps}
\end{figure*}

\subsection{Numerical experiments}

Figure~\ref{schematic} (not to scale) shows the setup of our numerical experiments. Two {\color{black} in-phase} walkers are initially placed at distances $r_1$ and $r_2$ from the origin and at equal angles from the $y$-axis. If the droplets were traveling at constant speed, the difference $\tau = |r_1-r_2|$ would be {\color{black}proportional} to the difference in their time of arrival at the origin. {\color{black} To limit the size of the parameter space to be explored, we have fixed the angle between the droplets' trajectories to $90^{\circ}$. However, we note that the detailed dynamics of the droplets are quite sensitive to the choice of this initial impact angle.} In the numerical simulations, the droplets are point particles and the blue disk indicates the non-dimensional size $\hat{d}=[0.13,0.21]$ corresponding to the diameters in the range $d=[0.6,1]$ mm of the physical droplet. The visible size of the wavepacket produced by the droplet's impact is a few times $\lambda_{\rm F}$. We assume an initial uncertainty in the center-of-mass of the droplets' positions (white dots in Fig.\ \ref{schematic}). This is modeled by drawing random numbers from a uniform distribution within a disk of the same diameter as a droplet, $\hat{d}$. The droplets are launched at constant speed $V$, corresponding to the stable walking speed of an isolated droplet given by the stroboscopic model for a single walker,\citep{Oza2013} and are directed towards the origin. The detection occurs at time $t=250$, during which a droplet traveling at speed $V$ would travel a distance of at least $2r_1$. {\color{black} This cut-off point of $t=250$ has been chosen to ensure that any transient dynamics of droplet pairs have decayed and the droplets have settled either into a stable uncorrelated state or one of the correlated bound states. We numerically integrate the equations of motion in \eqref{eq_1} using the modified Euler method described by Valani and Slim.\citep{twodroplets}} 

We vary the proximity to the Faraday threshold, $\gamma/\gamma_F$, which traces out a curve for a droplet of fixed diameter in the $\beta$-$\kappa$ parameter space. Figure~\ref{ps} shows two sets of such curves in the $\beta$-$\kappa$ parameter space. Each curve corresponds to a fixed value of the dimensionless drag coefficient $C$ and the {\color{black}dimensionless} droplet diameter {\color{black}$\hat{d}$}. Increasing $\gamma/\gamma_F$ increases $\beta$ and decreases $\kappa$.

\section{Results}

Two walkers are launched towards each other with varying path differences in the range {\color{black}$0\leq \tau \leq 18/2\pi$}. This is done by fixing  the distance {\color{black}$r_1=100/2\pi$} of the first droplet and changing the distance of the second droplet $r_2$ in the range {\color{black}$[82,100]/2\pi$}. {\color{black} Each droplet is set to have an uncertainty in the initial position of its center of mass. The droplets' initial positions are thus determined by drawing random numbers from a uniform distribution within a disk of diameter $\hat{d}$.} We observe four generic behaviours: (i) In the majority of cases the droplets remain uncorrelated and travel along straight lines in different directions. (ii) The droplets pair up into {\color{black}promenading} walkers where the droplets are walking parallel with sideways oscillations. More exotic {\color{black}promenading} walkers also arise as shown in the parameter space study of two walkers by Valani and Slim.\citep{twodroplets} We refer to {\color{black}all of} these as {\color{black}promenading} correlations. (iii) The droplets form a two-droplet orbiting pair referred to as orbiting correlations. (iv) The droplets pair up in a chasing mode where they are walking one behind the other. These are referred to as chasing correlations. 

\begin{figure}
\centering
\includegraphics[width=1\columnwidth]{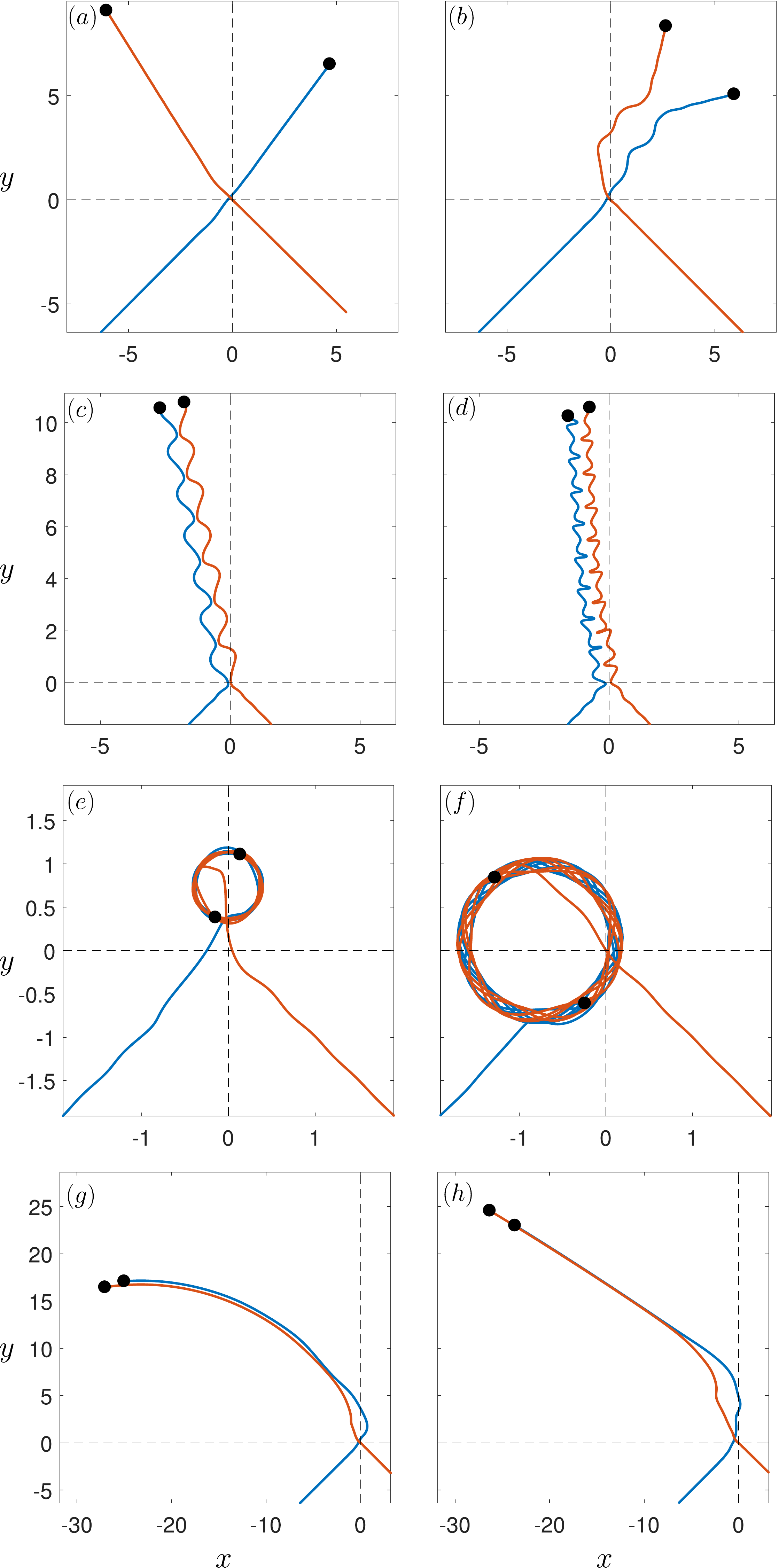}
\caption{Typical two-droplet trajectories: {\color{black}Uncorrelated trajectories whose inter-droplet distance diverges with time are shown in (a) and (b) with parameter values same as Fig.~\ref{Dip}(b) and path differences $\tau=2.71$ and $\tau=1.51$ respectively. Promenading correlations with symmetrical oscillations (c) at $\tau=0.40$ and other parameters same as Fig.~\ref{Dip}(f), and asymmetrical oscillations (d) at $\tau=0.16$ and other parameters same as Fig.~\ref{Dip}(e). Orbiting correlations of smaller diameter (e) at $\tau=0.95$ and larger diameter (f) at $\tau=1.99$ with other parameters same as Fig.~\ref{Dip}(c). Chasing correlations with a circular path (g) at $\tau=2.71$ and other parameters same as Fig.~\ref{Dip}(f), and on a straight line (h) at $\tau=1.25$ and other parameters same as Fig.~\ref{Dip}(d). The axis are in units of the Faraday wavelength $\lambda_F$.}}  
\label{traj}
\end{figure}

Typical realizations of the two droplet bound states and {\color{black}uncorrelated} trajectories are summarized in Fig.~\ref{traj}. For the uncorrelated trajectories, two walkers may travel straight through the interaction region as shown in Fig.~\ref{traj}(a), or unbind after a short interaction Fig.~\ref{traj}(b). Two walkers can also reflect off each other and end up uncorrelated. The {\color{black}first} type of bound state we observe are \emph{{\color{black}promenading} walkers}. We observe two generic types of oscillating modes for this state: a symmetrical mode, Fig.~\ref{traj}(c), in which walkers perform symmetric sideways oscillations while walking in parallel and their center-of-mass follows a straight line, and an asymmetric mode, Fig.~\ref{traj}(d), in which oscillations are asymmetric and the center-of-mass oscillates.  {\color{black} The second} type of bound state we observe in simulations are \emph{orbiting walkers}. In this state the two walkers orbit around their common center of mass. We observe two types of orbiting correlations: In the first type, the droplets are orbiting at a closer inter-droplet distance  {\color{black}of $d^o_{12}\approx 0.8$}, Fig.~\ref{traj}(e), while in the second type they are orbiting at a larger inter-droplet distance  {\color{black}of $d^o_{12}\approx 1.8$} Fig.~\ref{traj}(f). Droplets orbiting at a larger distance are also found to sometimes oscillate radially. The third kind of correlation we observe are \emph{chasers}. In this state, the droplets are chasing one after another and are traveling at a constant speed and a fixed inter-droplet distance. We classify the chasers into two types: those which follow a circular path, Fig.~\ref{traj}(g), and those which walk on a straight line, Fig.~\ref{traj}(h). Straight-line chasers are found to have inter-droplet distances of {\color{black}$d^c_{12}\approx 3$} and {\color{black}$d^c_{12}\approx4$}, while the chasers which follow a circular path are closer with distances {\color{black}$d^c_{12}\approx1.3$} and {\color{black}$d^c_{12}\approx2$}. Moreover, the trailing walker is offset sideway from the leading walker in this latter case. {\color{black} Chasers have previously been identified in a bounded domain where walkers were confined into an annular region.\citep{PhysRevE.92.041004} The quantization and stability of chasers in an unbounded domain was studied numerically by Durey and Milewski.\citep{durey_milewski_2017} We have also observed straight line chasers at relatively high memory using a simple experimental apparatus when walkers are confined in a circular bath.}

\begin{figure*}
\centering
\includegraphics[width=2\columnwidth]{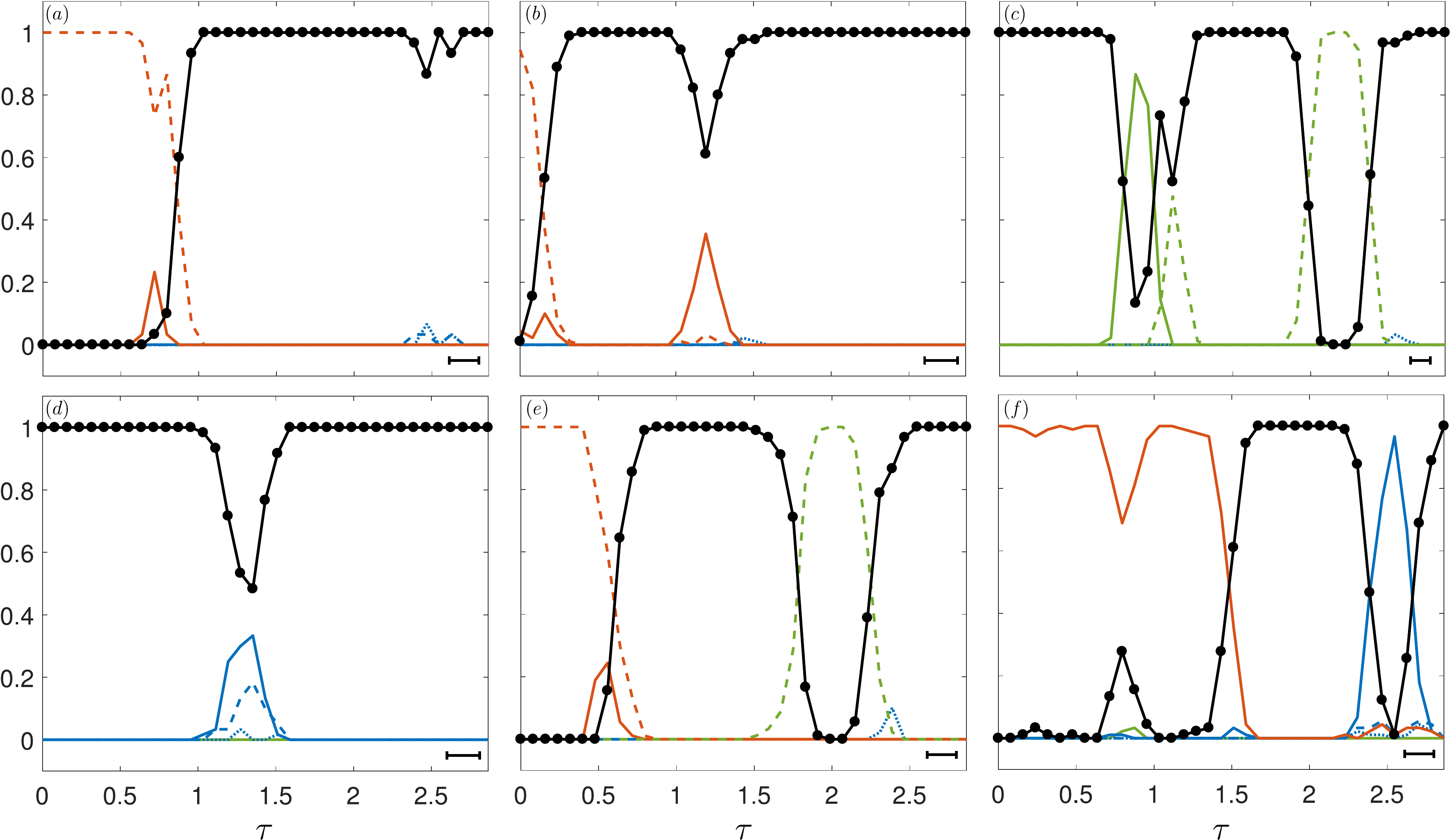}
\caption{Probability of {\color{black}different correlations} as a function of the path difference $\tau$ (in units of Faraday wavelength $\lambda_F$). The six plots {\color{black}$(a)$-$(f)$} correspond to the six points A-F in the $\beta$-$\kappa$ parameter space plot in {\color{black}Fig.~\ref{ps}}. The parameter values for these are, {\color{black}$(a)$} {\color{black} $\hat{d}=0.19$} and $\gamma/\gamma_F=0.86$, {\color{black}$(b)$}  {\color{black} $\hat{d}=0.21$} and $\gamma/\gamma_F=0.77$,  {\color{black}$(c)$} {\color{black} $\hat{d}=0.13$} and $\gamma/\gamma_F=0.88$, {\color{black}$(d)$}  {\color{black} $\hat{d}=0.21$} and $\gamma/\gamma_F=0.88$, {\color{black}$(e)$}  {\color{black} $\hat{d}=0.19$} and $\gamma/\gamma_F=0.85$ and  {\color{black}$(f)$} {\color{black} $\hat{d}=0.19$}, $\gamma/\gamma_F=0.9$. Thick black lines indicate the  {\color{black}probability of uncorrelated walkers} while the coloured lines show {\color{black}probabilities of} each of the different types of correlations: the symmetrical (red solid line)  and asymmetrical (red dashed line) modes of   {\color{black}promenading} walker correlations, smaller (green solid line) and larger (green dashed line) diameter orbiting correlations, circular-path chasing correlations with inter-droplet distance {\color{black}$d_{12}\approx2$} (blue solid line) and straight-line chasing correlations with inter-droplet distance {\color{black}$d_{12}\approx3$} (blue dashed line) and {\color{black}$d_{12}\approx4$} (blue dotted line). {\color{black}Each data point has a statistical uncertainty of $1/\sqrt{90}$ since each data point is obtained by averaging over 90 trajectories with slightly different initial conditions due to uncertainty in the initial positions of the droplets resulting from their non-zero size.} The horizontal black bar indicates the diameter of {\color{black}each} droplet.}
\label{Dip}
\end{figure*}

To understand the probabilistic properties of the emergence of two-walker correlations, we have studied the statistics of the two-droplet bound states by simulating many trajectories with similar initial conditions. On studying the final state of the droplets as functions of the path difference $\tau$ we find regions of correlated and uncorrelated behaviours. The parameter space chosen for the results presented is summarized in Fig.~\ref{ps}. The light gray region in the background corresponds to where initially parallel walkers unbind at late times in simulations with $\hat{\alpha}=0$, while white regions correspond to where initially parallel walkers remain in a bound state at late times (see \citet{twodroplets}). We primarily observe orbiting and chasing correlations and uncorrelated behavior for parameters in the light gray region, consistent with   {\color{black}promenading} walkers being unstable there. However, we do {\color{black}sometimes} observe   {\color{black}promenading} walkers near the edges of the light gray region, which may be because the simulations here are relatively short or may result from $\hat{\alpha}\ne0$. {\color{black} On traversing one of the curves corresponding to $C=0.17$ (low drag), we find that the low memory region (small $\beta$) is dominated by orbiting correlations with the emergence of promenading correlations at mid-memory and then only chasing correlations at high memory (large $\beta$). On the other hand, traversing a curve for $C=0.33$ (high drag), we find both promenading and orbiting correlations at low memory, promenading and chasing correlations at mid-memory and only chasing correlations at high memory. At even higher memories corresponding to $\beta>20$, we either observe chasing correlations or uncorrelated walkers.}

Figure~\ref{Dip} shows the probabilities {\color{black} of droplets being correlated or} uncorrelated at $t=250$ as functions of the path difference $\tau$. If the walkers end up in any of the droplet {\color{black} bound} states at the end of the simulation, we classify them as correlated, while if the droplets are found to be separated by a distance exceeding $30$ {\color{black} and moving in different directions}, we classify them as uncorrelated. The thick black solid lines in {\color{black}Fig.~\ref{Dip}} {\color{black}shows the probability of uncorrelated walkers while the coloured lines show the probabilities of each different types of correlations}. Every dip in the {\color{black}black} curves corresponds to a certain type of two-droplet correlation. Each of the data points comprises an ensemble average over $90$ simulated trajectories. The parameter values used for obtaining the data in Fig.~\ref{Dip} are indicated in Fig.~\ref{ps} as A-F. In Fig.~\ref{Dip}{\color{black}(a)}, we observe a single dominant dip from the asymmetrical mode of {\color{black}promenading} correlations (red dashed line) for {\color{black}$0<\tau\lesssim 1$}. {\color{black}Figures~\ref{Dip}(b) and (c)} comprise of two dips. {\color{black}Figure~\ref{Dip}(b)} has two dips arising from the asymmetrical and symmetrical modes of   {\color{black}promenading} walkers correlation at $\tau \approx 0$ and {\color{black}$\tau\approx 1.2$}{\color{black},} respectively. The two dips in {\color{black}Fig.~\ref{Dip}(c)} are from the smaller (green solid line) and the larger (green dashed line) diameter orbiting correlations. The dip near {\color{black}$\tau\approx 1$} comprises of both types of orbiting correlations while the dip near {\color{black}$\tau \approx 2.2$} is only from the larger diameter orbiting correlation. {\color{black}Figure~\ref{Dip}(d)} has a single dip near {\color{black}$\tau\approx 1.3$} but it is {\color{black} dominated by} a mixture of both straight-line (blue dashed line) chasing correlations with inter-droplet distance {\color{black}$d^c_{12}\approx3$} as well as circular-path (blue solid line) chasing correlations with inter-droplet distance {\color{black}$d^c_{12}\approx2$}. {\color{black}Figures~\ref{Dip}(e) and (f)} each have two dips arising from two different types of correlations. Dips in {\color{black}Fig.~\ref{Dip}(e)} arise from the asymmetrical mode of   {\color{black}promenading} walker correlations and larger diameter orbiting correlations, while the dips in {\color{black}Fig.~\ref{Dip}(f)} are from the symmetrical mode of {\color{black}promenading} walker correlations and circular path chasing correlations.

\section{Conclusions}

We have considered the non-Markovian dynamics of pairs of walking droplets with crossing paths that are initially separated by a large inter-droplet distance. We have studied the probability that the droplets remain unbound as a function of their path difference to the common origin. We have found three generic classes of two-droplet correlations: {\color{black}promenading}, orbiting and chasing, that are identified as dips in these graphs.

Our numerical experiments correspond to a hydrodynamic analog of the Hong--Ou--Mandel (HOM) two-photon interference experiment without a beam splitter. {\color{black} One of the limitations of the hydrodynamic pilot-wave model used in this study is the assumption of a constant impact phase. If one were to study these phenomena experimentally, the modulations of the impact phase may occur and may result in either enhancement or suppression of correlation dips. Another feature we observe here is that the correlation dips are very sensitive to the system parameters and an experimental realization of this setup may result in quantitatively different correlation dips but we do expect the qualitative features of the correlation dips to persist.}

It is known that for both classical waves and for classical particles that the visibility of the HOM dip cannot exceed 50\% .\citep{PhysRevLett.96.240502} It is therefore reasonable to assume the same to be true also for classical composite objects comprised of a droplet \emph{and} a wave. 
To achieve a closer analog of the HOM interference experiment, a hydrodynamic equivalent of a 50/50 beam splitter would need to be implemented. One such candidate could be a subsurface barrier with which a hydrodynamic analog of quantum tunneling has been demonstrated.\citep{PhysRevLett.102.240401,PhysRevFluids.2.034801,PhysRevE.95.062607} If the height and the width of such a barrier are suitably tuned, then in principle it should be possible to have a subsurface barrier reflect or transmit a single walker with a 50\% probability. However, the reflection of a walker from a subsurface barrier is known to be sensitive to the incident angle and it might be difficult to overcome this subtlety in practice. Another scenario, motivated by the atomic HOM experiments,\cite{Lopes2015} would be to direct the droplets through a grid barrier that would act as a Bragg diffraction grating for the droplets. An \textit{ad hoc} numerical implementation of a beam splitter can be realized by reversing the $x$-component of the droplets' velocity along with their memories with a 50\% probability once the droplets enter a spatial `beam splitter region'. On testing this idea, we found that the qualitative features of the two-droplet correlation dips persist.
In summary, we have demonstrated a richness in the reaction dynamics of two walkers paving the way to further studies of many-droplet correlated behaviors of these curious non-Markovian dynamical systems.

\section*{Acknowledgements}
R. Valani is grateful for discussions with J. W. M. Bush. This research was funded by an Australian Government Research Training Program (RTP) Scholarship to R.V. 
\section*{References}

%\nocite{*}
\bibliography{HOM_revised}% Produces the bibliography via BibTeX.

\end{document}